\documentclass[aps,prx,reprint,twocolumn, amsmath,amssymb,amsfonts,footinbib,
superscriptaddress,longbibliography,
notitlepage]{revtex4-2}

\usepackage[T1]{fontenc}
\usepackage[utf8]{inputenc}

\usepackage{amsmath}
\usepackage{amssymb}
\usepackage{mathtools}
\usepackage{graphicx}
\usepackage[percent]{overpic}
\usepackage{mathrsfs}
\usepackage{bm}
\usepackage{braket}
\usepackage{booktabs}
\usepackage[version=4]{mhchem}
\usepackage{xcolor}
\usepackage{soul}
\usepackage{comment}

\usepackage{hyperref}

\setstcolor{red}

\definecolor{lightblue}{rgb}{.90,.95,1}
\definecolor{greeny}{rgb}{0.9,1,0}
\sethlcolor{greeny}
\definecolor{jaune}{RGB}{255,253,55}
\sethlcolor{jaune}

\definecolor{cred}{RGB}{228,26,28}
\definecolor{cblue}{RGB}{8,48,107}
\definecolor{cgreen}{RGB}{77,175,74}
\definecolor{cgray}{RGB}{150,150,150}
\definecolor{clgray}{RGB}{200,200,200}
\definecolor{cpurple}{RGB}{152,78,163}
\definecolor{corange}{RGB}{255,127,0}
\definecolor{cgold}{RGB}{230,171,2}

\definecolor{cred}{RGB}{188,55,84}
\definecolor{cblue}{RGB}{55,126,184}
\hypersetup{colorlinks=true,linkcolor=cred,citecolor=cred,urlcolor=blue}

\makeatletter
\renewcommand{\fnum@figure}{Fig.~\thefigure}
\makeatother

\graphicspath{{./figs/}}

\begin{document}
\title{Relevant--marginal field mixing resolves the four-state Potts endpoint in the square-lattice $J_1$--$J_2$ Ising model}

\author{Yihua Sun}
\affiliation{Research Center for Quantum Physics and Technologies, Inner Mongolia University, Hohhot 010021, China}
\affiliation{School of Physical Science and Technology, Inner Mongolia University, Hohhot 010021, China}

\author{Yuchen Fan}
\email{yuchenfan@imu.edu.cn}
\affiliation{Research Center for Quantum Physics and Technologies, Inner Mongolia University, Hohhot 010021, China}
\affiliation{School of Physical Science and Technology, Inner Mongolia University, Hohhot 010021, China}

\begin{abstract}
Identifying endpoint criticality is challenging because pseudo-first-order signatures can persist over accessible system sizes and obscure the underlying scaling behavior. This difficulty is especially acute in the frustrated square-lattice \(J_1\)--\(J_2\) Ising model, where the location and even the nature of the stripe-ordering endpoint have remained controversial for decades. Here we treat this endpoint as a relevant--marginal two-field scaling problem and develop a generalized field-mixing construction to resolve the endpoint ambiguity. Mixed-histogram symmetry defines a finite-size pseudotransition manifold, transforming pseudo-first-order finite-size structure into a scaling trajectory on which the normal relevant detuning is removed and the residual marginal drift is exposed. Along this manifold, Binder analysis locates the endpoint as a distinct singular point, while mixed-histogram scaling, comparison with the Ashkin--Teller Potts point, and complementary Binder scaling establish four-state Potts endpoint criticality with logarithmic corrections.
\end{abstract}

\date{\today}
\maketitle

{\it{Introduction.}}---
Critical endpoints occupy a special place in the theory of phase transitions: they are singular points where a continuous transition line terminates on a first-order boundary. Unlike an ordinary point on a critical line, a critical endpoint is specified by multiple local scaling fields, including a normal field that tunes across the phase boundary and a tangent field that selects the endpoint along it~\cite{Cardy1996}. Their numerical identification is therefore intrinsically delicate. On one side of the endpoint, the transition is genuinely discontinuous and exhibits phase coexistence; on the other side, the transition is continuous, but finite systems can still retain coexistence-like, pseudo-first-order signatures.

A powerful way to implement such a local scaling analysis is field mixing, which replaces the bare thermodynamic parameters by mixed scaling fields adapted to the singular point. This idea has been particularly successful in endpoint and multicritical problems governed by two relevant scaling directions~\cite{Bruce1992,Wilding1992,Wilding1996,Plascak2013,Kwak2015,Mataragkas2023,wlzf-xy92}.
A paradigmatic example is the Blume--Capel model, whose tricritical Ising endpoint can be identified by constructing mixed energy-like operators and imposing symmetry of the corresponding probability distribution~\cite{Kwak2015,Mataragkas2023}. In such cases, field mixing converts an otherwise ambiguous finite-size problem into a scaling analysis in the proper local coordinates of the singular point.

A natural arena for this endpoint-scaling perspective is the long-standing endpoint problem in the frustrated square-lattice \(J_1\)--\(J_2\) Ising model~\cite{NIGHTINGALE1977486,Barber1979,PhysRevLett.43.177,JOitmaa1981,PhysRevB.21.1941,PhysRevB.21.1285,PhysRevB.31.5946,FAguilera-Granja1993,PhysRevB.48.3519,JLMoran-Lopez1994,LOPEZSANDOVAL1999437,DOSANJOS20081180,PhysRevE.76.021123,Malakis2006,Kalz2008,AKalz2009,Murtazaev2011,PhysRevB.84.174407,PhysRevLett.108.045702,PhysRevB.87.144406,PhysRevB.86.134410,Murtazaev2013,PhysRevE.91.032145,MURTAZAEV20151,RAMAZANOV201635,PhysRevE.104.024118,PhysRevE.108.054124,PhysRevB.109.104419,PhysRevE.111.024109,PhysRevB.104.144429,PhysRevE.107.034124,PhysRevB.109.064422,10.1093/ptep/ptae061,PhysRevE.111.024131}. The stripe-ordered phase breaks a lattice \(Z_4\) symmetry, and the standard picture identifies the continuous stripe-ordering boundary with the Ashkin--Teller universality line, terminating at its low-temperature end in a four-state Potts endpoint. Yet both the location and the nature of this endpoint remain unsettled, with proposed scenarios including a finite four-state Potts endpoint~\cite{PhysRevLett.108.045702,PhysRevB.87.144406}, tricritical Ising endpoint~\cite{PhysRevE.104.024118}, and a weak-first-order boundary without a finite endpoint~\cite{PhysRevB.109.104419}.
Even within the standard Potts scenario, estimates of the endpoint location \(g_{\mathrm{ep}}\) have spread over a broad intermediate-frustration range~\cite{PhysRevB.48.3519,JLMoran-Lopez1994,LOPEZSANDOVAL1999437,DOSANJOS20081180,PhysRevE.76.021123,Malakis2006,Kalz2008,AKalz2009,Murtazaev2011,PhysRevB.84.174407,PhysRevLett.108.045702,PhysRevB.87.144406}. The difficulty arises from the coexistence of pseudo-first-order finite-size signatures with the logarithmically slow drift associated with the putative marginally irrelevant field of the Potts endpoint~\cite{PhysRevLett.44.837,Salas1997,PhysRevB.22.2560}. Despite extensive work, this endpoint has not been treated explicitly as a full relevant--marginal two-field endpoint-scaling problem that separates the normal relevant detuning from the tangent marginal drift.

Here we develop a generalized field-mixing framework for endpoint criticality governed by one relevant scaling field and a marginally irrelevant one, and apply it to the \(J_1\)--\(J_2\) Ising endpoint as a relevant--marginal two-field scaling problem. Central to the construction is a finite-size pseudotransition manifold that removes the normal relevant detuning and exposes the residual marginal drift. Binder analysis along this manifold then locates the endpoint as a distinct singular point, rather than relying on apparent Potts-like scaling along the phase boundary. For the \(J_1\)--\(J_2\) Ising endpoint, this endpoint-first construction yields direct histogram-level evidence for four-state Potts universality with logarithmic corrections. This identification is independently corroborated by comparison with the Potts point of the Ashkin--Teller model and by Binder scaling in both the magnetic and nematic sectors. Our results resolve a longstanding ambiguity in the \(J_1\)--\(J_2\) Ising model and establish a field-mixing route to relevant--marginal endpoint criticality.

\begin{figure}[t!]
\includegraphics[angle=0,width=0.84\linewidth]{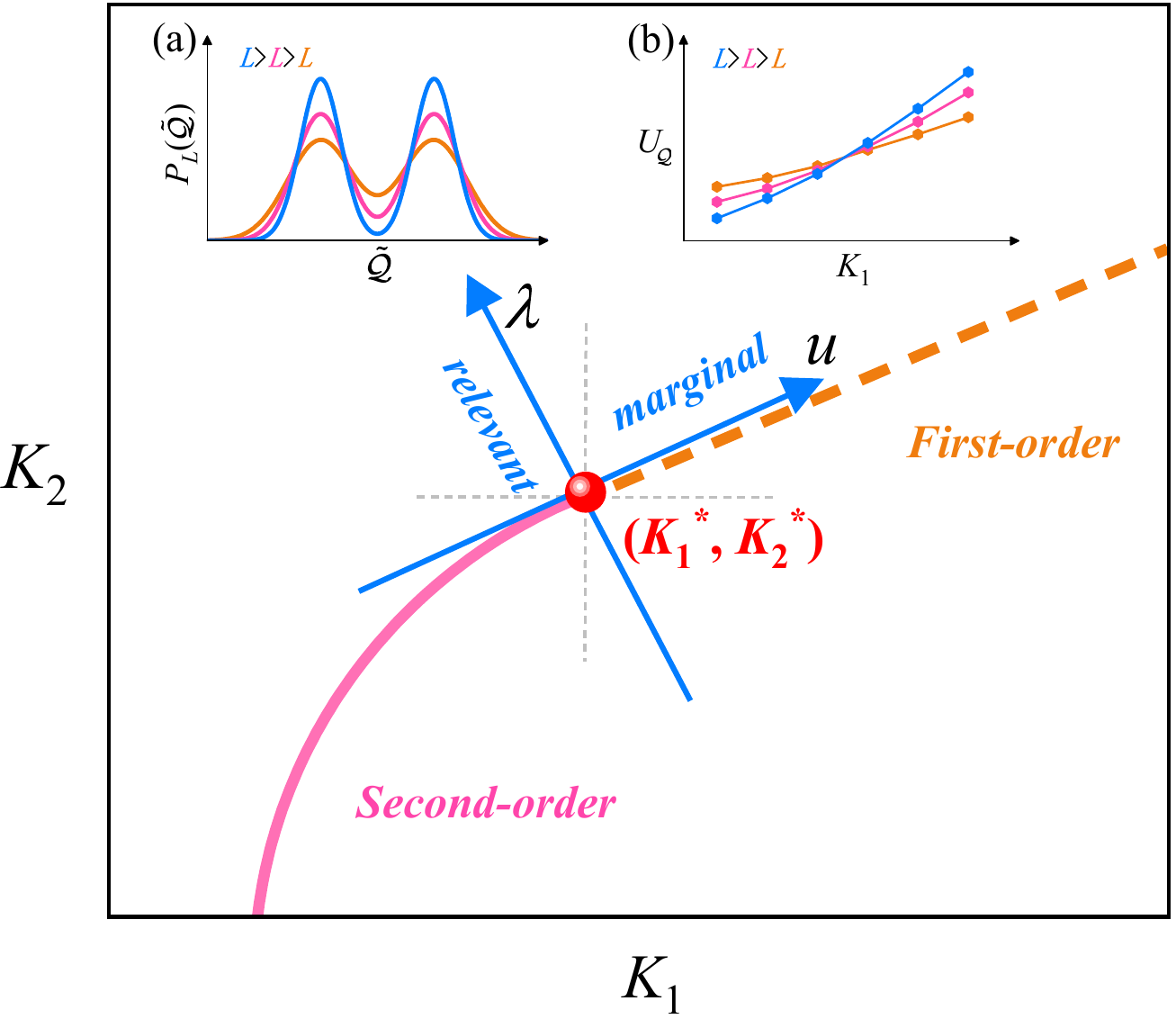}
\caption{
Generalized field mixing for endpoint criticality with marginal flow.
Schematic illustration of the generalized field-mixing framework near a critical endpoint in the $(K_1,K_2)$ phase diagram, where a first-order line terminates on a continuous transition line. Taking the endpoint as the origin, two mixed variables are introduced: $\lambda$ for the relevant detuning direction and $u$ for the tangent marginally irrelevant direction. This separation enables the independent analysis of the relevant critical singularity and the residual marginal drift. Inset (a), finite-size pseudotransition points near the endpoint are determined by requiring a symmetric histogram of the centered mixed variable \(\tilde{\mathcal Q}\). Inset (b), Binder analysis performed along the finite-size pseudotransition manifold locates the endpoint through the crossing points.
}
\label{fig:1}
\end{figure}

{\it{Generalized field mixing.}}--- 
We begin from the standard field-mixing construction for a Hamiltonian containing two energy-like sectors,
\begin{equation}
H = J_1 E_1 + J_2 E_2 ,
\label{eq:generic_two_term_H}
\end{equation}
where the natural bare control parameters are the dimensionless couplings
\(
K_1=\beta J_1
\)
and
\(
K_2=\beta J_2
\),
with \(\beta=1/T\) the inverse temperature (setting \(k_B=1\)). The corresponding intensive energy-like operators are
\(
\mathcal{E}_1\equiv L^{-d}E_1
\)
and
\(
\mathcal{E}_2\equiv L^{-d}E_2
\).
Near an endpoint \((K_1^\ast,K_2^\ast)\), the bare deviations
\(
\delta K_1=K_1-K_1^\ast
\)
and
\(
\delta K_2=K_2-K_2^\ast
\)
are reorganized along the local scaling directions as
\begin{equation}
\lambda = \delta K_2 + r\,\delta K_1,
\qquad
u = \delta K_1 + s\,\delta K_2,
\label{eq:lambda_u_generic}
\end{equation}
as illustrated in Fig.~\ref{fig:1}, with \(r\) and \(s\) denoting the corresponding mixing coefficients. Here \(\lambda\) parametrizes the generic normal detuning direction away from the phase boundary and therefore represents the relevant scaling field, whereas \(u\) runs locally tangent to the phase boundary and, at the endpoint, corresponds to the marginally irrelevant direction. The associated scaling operators are
\begin{equation}
\mathcal{Q} = \frac{1}{1-rs}\left(\mathcal{E}_2-s\,\mathcal{E}_1\right),
\quad
\mathcal{E} = \frac{1}{1-rs}\left(\mathcal{E}_1-r\,\mathcal{E}_2\right),
\label{eq:Q_E_generic}
\end{equation}
satisfying
\(
\langle\mathcal{Q}\rangle=-L^{-d}\partial_{\lambda}\ln Z
\)
and
\(
\langle\mathcal{E}\rangle=-L^{-d}\partial_{u}\ln Z
\).

Conventional field mixing assumes two relevant directions~\cite{Kwak2015,Mataragkas2023}. Endpoint criticality with a marginally irrelevant flow is qualitatively different and therefore requires a modified finite-size scaling description. At the endpoint, the mixed scaling fields are tuned to \(\lambda=0\) and \(u=0\), and the joint histogram is governed by the corresponding scaling operators together with the logarithmic correction induced by the marginally irrelevant field.
Its joint distribution is therefore expected to obey the scaling form
\begin{equation}
P_L \propto
\mathcal{P}^{\ast}\!\left(
L^{d-y_t}(\ln L)^{-q}\,\tilde{\mathcal{Q}},\;
X_u(\mathcal E;L)
\right),
\label{eq:joint_endpoint_hist_prop}
\end{equation}
where \(\mathcal{P}^{\ast}\) is a universal scaling function, \(y_t\) is the RG eigenvalue of the relevant direction, and \(q\) denotes the corresponding logarithmic correction exponent.
Here \(\tilde{\mathcal Q}\equiv \mathcal Q-\langle\mathcal Q\rangle\) denotes the centered fluctuation of the scaling operator \(\mathcal Q\) conjugate to the relevant field \(\lambda\). Its critical scale at the endpoint is \(L^{y_t-d}(\ln L)^q\), making the first argument the dimensionless scaling variable for the relevant direction, while \(X_u(\mathcal E;L)\) represents the residual tangent variable associated with the marginally irrelevant sector.
Projecting the joint distribution onto \(\tilde{\mathcal Q}\), that is, integrating out the remaining scaling direction, then gives the scaling form
\begin{equation}
P_L(\tilde{\mathcal{Q}})
=
L^{d-y_t}(\ln L)^{-q}
\,\mathcal{P}_\mathcal Q^{\ast}\!\left[
L^{d-y_t}(\ln L)^{-q}\,
\tilde{\mathcal{Q}}
\right],
\label{eq:Q_endpoint_scaling}
\end{equation}
where \(\mathcal{P}_\mathcal Q^{\ast}\) is the associated universal scaling function, and the prefactor \(L^{d-y_t}(\ln L)^{-q}\) is fixed by the normalization of the probability density.

\begin{figure*}[t!]
\includegraphics[angle=0,width=1.0\linewidth,height=7.3cm]{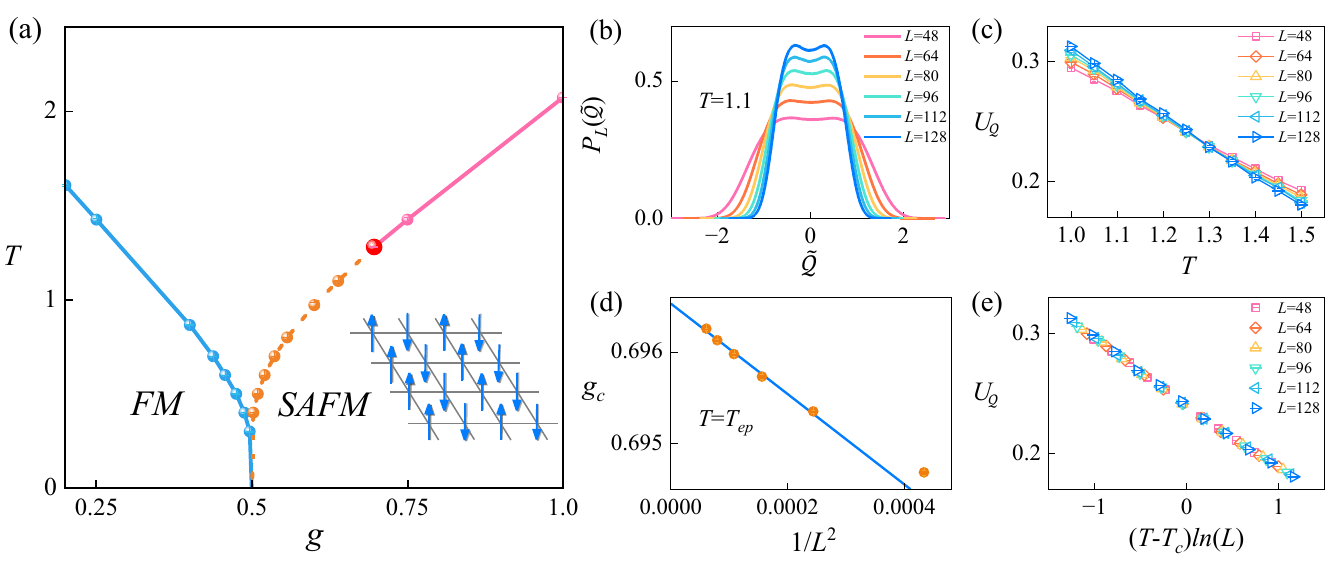}
\caption{
Location of the endpoint in the square-lattice \(J_1\)–\(J_2\) Ising model.
(a) Phase diagram of the square-lattice \(J_1\)–\(J_2\) Ising model. The transition from the SAFM phase to the paramagnetic phase is first order at low temperatures and continuous at higher temperatures, with the endpoint located at \((g_{\mathrm{ep}},T_{\mathrm{ep}})=(0.696(10),1.28(3))\).
(b) Maximally symmetric histograms \(P_L(\tilde{\mathcal{Q}})\) at \(T=1.1\) for different system sizes. The symmetry condition determines the finite-size pseudotransition point, at which the fourth-order cumulant \(U_\mathcal Q\) is evaluated.
(c) Fourth-order cumulant \(U_\mathcal Q\) measured along the finite-size pseudotransition manifold. The crossings for different system sizes provide an estimate of the endpoint temperature \(T_{\mathrm{ep}}=1.28(3)\). 
(d) Finite-size pseudotransition couplings \(g_c(L)\) at \(T=T_{\mathrm{ep}}=1.28\), extracted from the maximally symmetric endpoint histograms shown in Fig.~\ref{fig:3}(a). Extrapolating \(g_c(L)\) to the thermodynamic limit gives the conditional estimate \(g_{\mathrm{ep}}|_{T=1.28}=0.6965(3)\).
The quoted uncertainty is obtained from the fixed-$T$ finite-size extrapolation, with the statistical uncertainties of the histogram-determined $g_c(L)$ values included. The uncertainty in $g_{\mathrm{ep}}=0.696(10)$ additionally includes the uncertainty in $T_{\mathrm{ep}}$.
(e) Finite-size scaling analysis of \(U_\mathcal Q\) near the endpoint with logarithmic finite-size dependence.
}
\label{fig:2}
\end{figure*}

To locate the endpoint, we analyze the histogram of \(\tilde{\mathcal Q}\), which provides a direct probe of coexistence-like finite-size behavior.
For each system size, we implement this construction by fixing one bare parameter (e.g., \(K_2\)) and tuning the other control parameter together with the mixing coefficient \(s\) so that \(P_L(\tilde{\mathcal{Q}})\) is maximally symmetric (as schematically illustrated in inset (a) of Fig.~\ref{fig:1}).
On the first-order side, this reduces to the usual equal-weight condition; beyond the endpoint, it extends smoothly into the pseudo-first-order regime.
The symmetric-histogram points thus define the finite-size pseudotransition manifold along which the normal relevant detuning is tuned to zero.
Along this manifold, we evaluate the fourth-order cumulant
\begin{equation}
U_\mathcal Q(L)=1-\frac{\langle \tilde{\mathcal{Q}}^4\rangle}{3\langle \tilde{\mathcal{Q}}^2\rangle^2},
\label{eq:UQ_def_main}
\end{equation}
and use its finite-size crossings to locate the endpoint (as schematically illustrated in inset (b) of Fig.~\ref{fig:1}).
With the relevant detuning removed, the leading size dependence of \(U_\mathcal Q\) is governed by the marginally irrelevant sector.

\begin{figure}[t]
\includegraphics[width=1.0\linewidth,height=7.3cm]{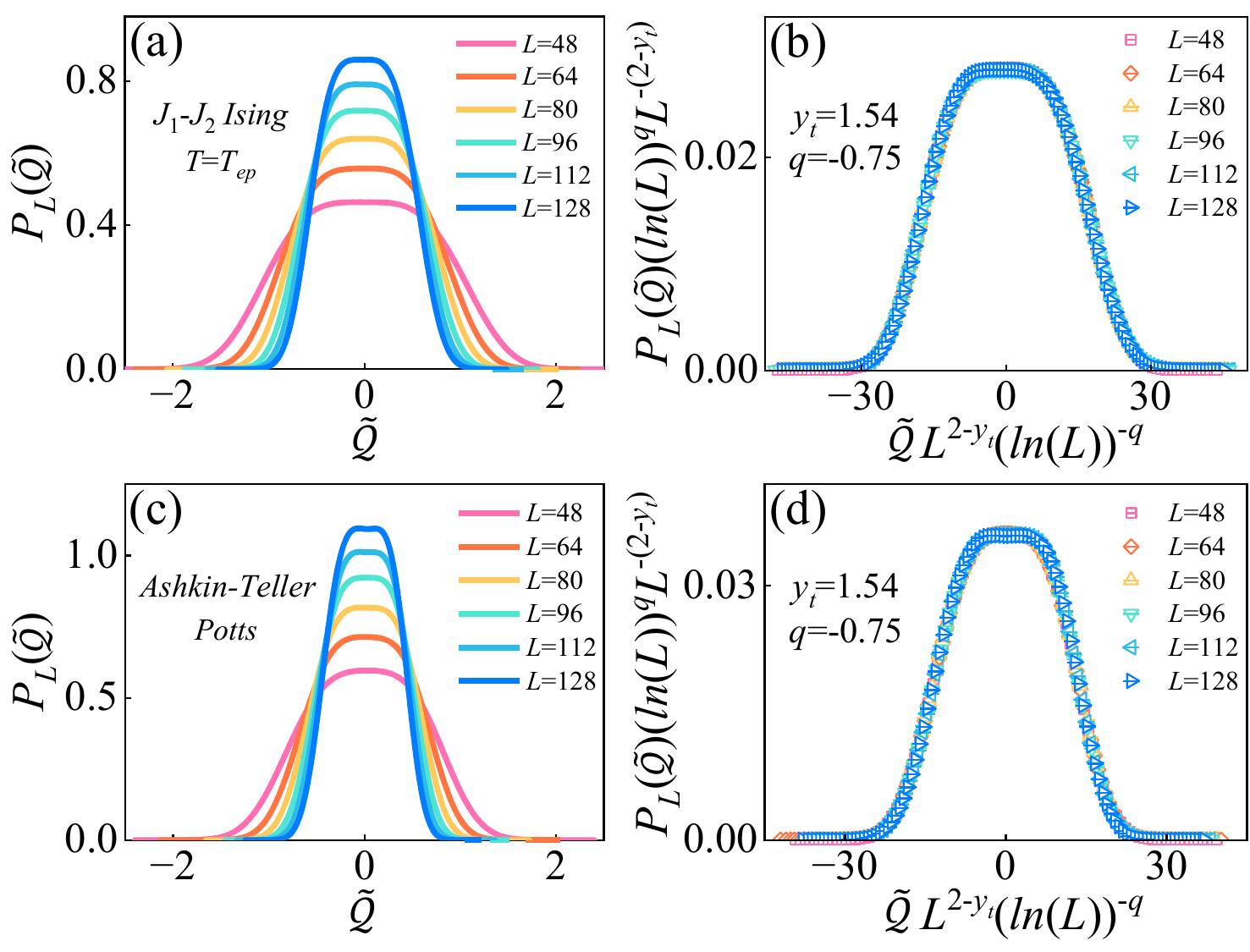}
\caption{
Histogram scaling at the endpoint and comparison with the Ashkin--Teller Potts point.
(a) Maximally symmetric histograms \(P_L(\tilde{\mathcal{Q}})\) at \(T=T_{\mathrm{ep}}\) of the square-lattice \(J_1\)–\(J_2\) Ising model for different system sizes.
(b) Finite-size scaling collapse of the endpoint histograms shown in (a), using the logarithmically corrected four-state Potts histogram scaling form.
(c) Maximally symmetric histograms \(P_L(\tilde{\mathcal{Q}})\) at the four-state Potts point of the Ashkin--Teller model for different system sizes.
(d) Logarithmically corrected finite-size scaling collapse of the histograms in (c).
}
\label{fig:3}
\end{figure}

{\it{The square-lattice \(J_1\)--\(J_2\) Ising model and endpoint identification.}}---  
We now apply the generalized field-mixing framework to the frustrated square-lattice \(J_1\)--\(J_2\) Ising model, with Hamiltonian
\begin{equation}
H=J_1\sum_{\langle ij\rangle}\sigma_i\sigma_j
+J_2\sum_{\langle\langle ij\rangle\rangle}\sigma_i\sigma_j,
\label{eq:J1J2_H}
\end{equation}
where \(\sigma_i=\pm1\) are Ising spins. We take \(J_1<0\) and \(J_2>0\), corresponding to ferromagnetic nearest-neighbor and antiferromagnetic next-nearest-neighbor couplings. This convention is equivalent, under a staggered gauge transformation, to the choice \(J_1>0\), \(J_2>0\). Throughout this work we set \(|J_1|=1\) and parametrize frustration by \(g\equiv J_2/|J_1|\).

We perform Monte Carlo simulations on \(L\times L\) square lattices with periodic boundary conditions, using system sizes up to \(L=128\). For the mixed-histogram analysis, long Markov chains are used to accumulate \(10^7\)--\(10^8\) samples after thermalization, while standard observables are averaged over 56--112 independent chains; convergence is checked by the stability of both the histograms and \(U_{\mathcal Q}\) upon increasing statistics.
The computational cost is dominated by the Markov-chain sampling required to obtain converged joint histograms of $(\mathcal{E}_1,\mathcal{E}_2)$. For the local-update scheme used here, one Monte Carlo sweep requires $\mathcal{O}(L^2)$ spin-update attempts, giving a total cost of $\mathcal{O}(N_{\mathrm{MC}}L^2)$, where $N_{\mathrm{MC}}$ is the number of sweeps. The required $N_{\mathrm{MC}}$ increases with system size near the endpoint because of increasing autocorrelation times. Once these joint histograms have been accumulated, histogram reweighting and the optimization of $g$ and $s$ add comparatively little computational cost for the present two-field problem.

At zero temperature, the model has two ordered phases: a ferromagnetic (FM) phase at small \(g\) and a stripe antiferromagnetic (SAFM) phase for \(g>1/2\). Because the SAFM state breaks the lattice \(Z_4\) symmetry, the continuous part of its thermal phase boundary is expected to belong to the Ashkin--Teller (AT) universality class, terminating at a four-state Potts endpoint that separates the low-temperature first-order segment from the high-temperature continuous branch~\cite{PhysRevLett.108.045702,PhysRevB.87.144406}. This is precisely the setting in which endpoint criticality is governed by one relevant field together with a marginally irrelevant one~\cite{PhysRevLett.44.837,Salas1997,PhysRevB.22.2560}.
At the same time, the location and even the nature of this endpoint remain highly controversial: Monte Carlo works established a close correspondence with the Ashkin--Teller model and suggested a four-state Potts endpoint at $g_{\mathrm{ep}}\approx0.67$~\cite{PhysRevLett.108.045702,PhysRevB.87.144406}, while recent infinite-size tensor-network studies have put forward competing alternatives, including a tricritical Ising endpoint~\cite{PhysRevE.104.024118} and a weak-first-order boundary without a finite endpoint, with the apparent Potts-like behavior interpreted as pseudocriticality~\cite{PhysRevB.109.104419}.
Logarithmically slow scaling and pseudo-first-order signatures make this ambiguity difficult to resolve.

We now turn to the direct numerical identification of the endpoint, the first step of the generalized field-mixing analysis, as summarized in Fig.~\ref{fig:2}. Fig.~\ref{fig:2}(a) shows that the SAFM--paramagnetic transition is first order at low temperatures and continuous at higher temperatures, with the two branches meeting at \((g_{\mathrm{ep}},T_{\mathrm{ep}})=(0.696(10),1.28(3))\). Representative transition-point determinations away from the endpoint are shown in Figs. S1 and S2~\cite{SM}. As illustrated in Fig.~\ref{fig:2}(b), for each system size we construct the finite-size pseudotransition point at a fixed temperature by tuning \(g\) and \(s\) to impose maximal symmetry on the histogram of the centered mixed variable \(\tilde{\mathcal Q}\). This criterion extends smoothly into the pseudo-first-order regime and yields both the finite-size pseudotransition point and the corresponding fourth-order cumulant \(U_\mathcal Q\).
Evaluating \(U_{\mathcal Q}\) along this manifold [Fig.~\ref{fig:2}(c)] gives the endpoint temperature \(T_{\mathrm{ep}}\) from the crossings, while extrapolation of the associated finite-size pseudotransition couplings \(g_c(L)\) at \(T=T_{\mathrm{ep}}\) [Fig.~\ref{fig:2}(d)] yields the thermodynamic-limit endpoint coupling \(g_{\mathrm{ep}}\).
The finite-size pseudotransition couplings are extrapolated as $g_c(L)=g_c+aL^{-d}$ with $d=2$. This inverse-volume form is motivated by first-order finite-size scaling~\cite{PhysRevB.30.1477,KBinder1987} and is applied here in the finite-size pseudo-first-order regime. This extrapolation has also been used in field-mixing studies of the Blume--Capel tricritical point~\cite{Kwak2015,Mataragkas2023,wlzf-xy92}.
Fig.~\ref{fig:2}(e) provides the corresponding scaling test: once the normal relevant detuning is removed, the residual drift of \(U_\mathcal Q\) is not organized by the power-law behavior expected for a second relevant field, but instead follows an approximately logarithmic dependence.

{\it{Histogram scaling at the endpoint}.}--- 
We next examine the endpoint histogram and its finite-size scaling. Figure~\ref{fig:3}(a) shows the maximally symmetric histograms
\(P_L(\tilde{\mathcal{Q}})\) at the endpoint temperature \(T=T_{\mathrm{ep}}\). 
At accessible system sizes, these endpoint histograms still display a weakly developed, nearly flat two-peak profile, superficially similar to the coexistence histograms found along the first-order line. 
Since the first-order line terminates at the endpoint, these first-order-like features should be understood as finite-size pseudo-first-order effects rather than evidence of phase coexistence.
This interpretation is supported by earlier work on the same model, where pseudo-first-order behavior and the limited diagnostic value of first-order-like histogram features were explicitly discussed~\cite{PhysRevLett.108.045702,PhysRevB.87.144406,Sandviknote,PhysRevB.86.134410}. Related pseudo-first-order signatures have likewise been reported in the Baxter--Wu model, another system in the four-state Potts universality class~\cite{Schreiber2005}.

Within the generalized field-mixing framework, the endpoint histogram can be analyzed directly through controlled finite-size scaling. For the four-state Potts endpoint, a marginally irrelevant operator generates multiplicative logarithmic corrections~\cite{PhysRevLett.44.837,Salas1997,PhysRevB.22.2560}.
As a result, the relevant thermal scaling field enters the finite-size scaling variable in the form
\(
\lambda(L)\sim \lambda\,L^{y_t}(\ln L)^{q},
\)
with \(y_t=3/2\) and \(q=-3/4\)~\cite{Salas1997}.
The fluctuation scale of the conjugate scaling operator is therefore expected to behave as
\(
\tilde{\mathcal Q}\sim L^{y_t-d}(\ln L)^{q},
\)
thereby fixing the scaling exponents entering Eq.~\eqref{eq:Q_endpoint_scaling}.
As illustrated in Fig.~\ref{fig:3}(b), the endpoint histograms collapse according to the logarithmically corrected four-state Potts scaling form, providing a direct histogram-level test of the endpoint scaling.
In the histogram scaling collapses, $q=-3/4$ is fixed at its theoretical four-state Potts value, while optimization of $y_t$ gives $y_t=1.54$, close to the exact value $3/2$. This small deviation from the exact value may reflect subleading corrections beyond the leading scaling form~\cite{Salas1997}; a similar deviation is observed at the exactly known Ashkin--Teller Potts point in Fig.~3(d).

To make this identification concrete, we benchmark the \(J_1\)--\(J_2\) endpoint against the Ashkin--Teller (AT) model at the Potts point (for background on the AT model, see Refs.~\cite{Baxter1982,Nienhuis1987,DELFINO2004521}). We consider the standard Hamiltonian
\begin{equation}
H_{\rm AT}
=
-J\sum_{\langle ij\rangle}\bigl(\sigma_i\sigma_j+\tau_i\tau_j\bigr)
-K\sum_{\langle ij\rangle}\sigma_i\sigma_j\tau_i\tau_j,
\label{eq:AT_H_main}
\end{equation}
where \(\sigma_i,\tau_i=\pm1\) are two Ising variables on each site, and we set \(J=1\) for simplicity.
In the field-mixing description, this model naturally contains the same two energy-like sectors: the quadratic spin interaction
\(
E_1=\sum_{\langle ij\rangle}(\sigma_i\sigma_j+\tau_i\tau_j)
\)
and the four-spin interaction
\(
E_2=\sum_{\langle ij\rangle}\sigma_i\sigma_j\tau_i\tau_j,
\)
whose mixed combination defines the scaling operator \(\mathcal Q\).
At the Potts point \(K=1\) on the critical line, the histograms of \(\tilde{\mathcal Q}\) [Fig.~\ref{fig:3}(c)] exhibit finite-size structures similar to those of the \(J_1\)--\(J_2\) endpoint, and the same logarithmically corrected collapse is obtained in Fig.~\ref{fig:3}(d).
This benchmark validates the generalized field-mixing construction in a known relevant--marginal Potts setting and supports the identification of the \(J_1\)--\(J_2\) endpoint with four-state Potts criticality.

\begin{figure}[t]
\centering
\includegraphics[width=1.0\linewidth,height=7.3cm]{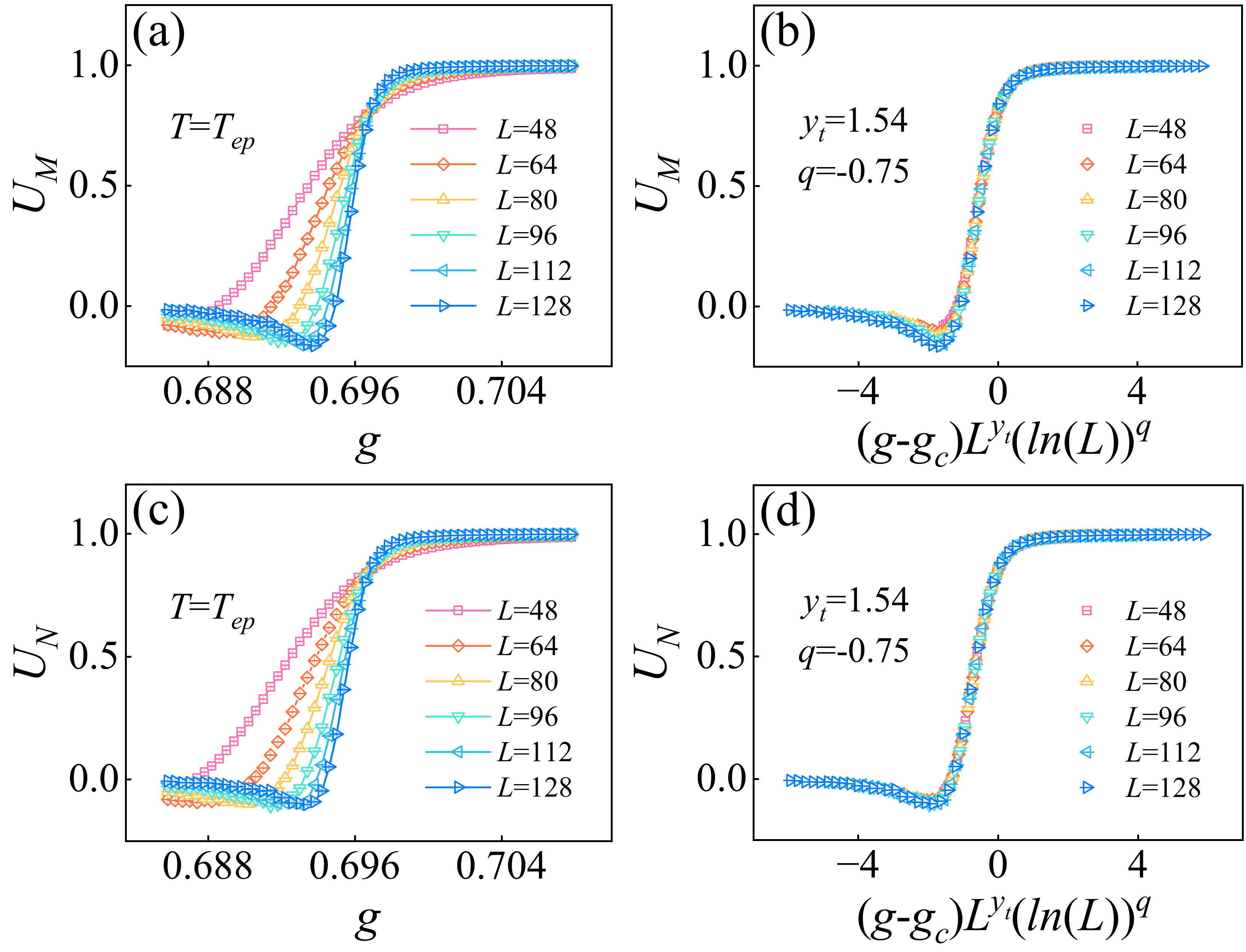}
\caption{
Binder scaling in the magnetic and nematic sectors.
(a) Binder cumulant of the stripe magnetic order parameter at \(T=T_{\mathrm{ep}}\) as a function of \(g\) for different system sizes, with crossings
at the endpoint coupling \(g_c=0.6968(2)\).
(b) Finite-size scaling collapse of the magnetic Binder data using the logarithmically corrected four-state Potts scaling form and the same exponents as in Fig.~3.
(c) Binder cumulant of the nematic order parameter at the same temperature, with crossings at the same endpoint coupling.
(d) Finite-size scaling collapse of the nematic Binder data using the same scaling form and exponents.
}
\label{fig:4}
\end{figure}

{\it{Independent confirmation from Binder scaling}.}---
With the endpoint identified by the field-mixing construction, we further test its critical structure through dimensionless Binder cumulants defined in complementary sectors of the stripe order. Using the standard definitions of Refs.~\cite{PhysRevLett.108.045702,PhysRevB.87.144406}, we first consider the two-component magnetic order parameter \(\mathbf m=(m_x,m_y)\), with
\(
m_x=\frac{1}{L^2}\sum_i \sigma_i (-1)^{x_i},
\quad
m_y=\frac{1}{L^2}\sum_i \sigma_i (-1)^{y_i},
\)
where \(L^2\) is the number of sites, \((x_i,y_i)\) are the integer lattice coordinates of site \(i\), and \(m^2=m_x^2+m_y^2\). The associated Binder cumulant
\begin{equation}
U_M=2\left(1-\frac{1}{2}\frac{\langle m^4\rangle}{\langle m^2\rangle^2}\right)
\label{eq:binder_def_main}
\end{equation}
probes the magnetic sector of the stripe order. To resolve the orientational component of the same \(Z_4\) stripe order, we further introduce the bond-nematic order parameter
\(
N=\frac{1}{L^2}\sum_i\left(\sigma_i\sigma_{i+\hat x}-\sigma_i\sigma_{i+\hat y}\right),
\)
where \(\hat x\) and \(\hat y\) denote the nearest-neighbor unit vectors along the two lattice directions. Its Binder cumulant is defined as
\begin{equation}
U_N=\frac{3}{2}\left(1-\frac{\langle N^4\rangle}{3\langle N^2\rangle^2}\right).
\label{eq:nematic_binder_def_main}
\end{equation}

Near the endpoint, we analyze Binder cumulants along the fixed-\(T\) cut \(T=T_{\mathrm{ep}}\). Since this cut generically projects onto the relevant scaling direction, their finite-size scaling takes the form~\cite{Salas1997},
\begin{equation}
U(g,L)=\mathcal{U}\!\left[(g-g_{\mathrm{ep}})L^{y_t}(\ln L)^{q}\right],
\label{eq:binder_scaling_main}
\end{equation}
with \(\mathcal{U}\) a universal scaling function. Figure~\ref{fig:4}(a) shows the magnetic Binder cumulant \(U_M\) versus \(g\) for different system sizes, with crossings converging to \(g_c=0.6968(2)\), in agreement with the endpoint position inferred from Fig.~\ref{fig:2}(d).
Although a negative Binder dip is commonly associated with phase coexistence and first-order behavior~\cite{PhysRevB.30.1477,Vollmayr1993}, the pronounced dip in \(U_M\) here reflects pseudo-first-order finite-size behavior near the Potts endpoint rather than a true first-order transition.
The coexistence-like signatures underlying the negative Binder dip can be visualized directly in the corresponding \(\mathbf m\) and \(m^2\) histograms (see Fig. S3~\cite{SM}). Including the logarithmic correction required by the four-state Potts RG structure allows the magnetic Binder data to collapse onto a single scaling curve [Fig.~\ref{fig:4}(b)]. Figures~\ref{fig:4}(c) and \ref{fig:4}(d) show that the nematic Binder cumulant \(U_N\) exhibits the same crossing point and the same logarithmically corrected scaling behavior, confirming that the identified Potts endpoint criticality is robust across complementary sectors of the underlying \(Z_4\) stripe order.

{\it{Discussion}.}---
In this work, we have recast the long-standing endpoint controversy in the \(J_1\)--\(J_2\) Ising model as a local scaling-field problem.
The difficulty in this model is the coexistence of pseudo-first-order finite-size signatures with the logarithmically slow flow of a marginally irrelevant field. The generalized field-mixing construction developed here resolves this ambiguity by defining a pseudotransition manifold that removes the normal relevant detuning and exposes the tangent marginal drift. This separation makes it possible to locate the endpoint and test its asymptotic criticality.

This endpoint-scaling perspective also clarifies how our results build on and extend earlier studies~\cite{PhysRevLett.108.045702,PhysRevB.87.144406}, which established the importance of pseudo-first-order behavior and provided strong evidence that the continuous branch terminates at a four-state Potts point. However, the endpoint itself was not isolated through an explicit endpoint-scaling construction, leaving room for competing interpretations. One distinguishing feature of the present analysis is its endpoint-first character: the endpoint is isolated by combining a mixed-histogram construction with Binder crossings along the finite-size pseudotransition manifold, rather than inferred indirectly from apparent Potts-like behavior along the phase boundary. In this construction, the first-order-like histogram structure is not merely an ambiguity to be avoided; its symmetry defines that manifold. The mixed histogram further provides a direct scaling test of the logarithmically corrected Potts form, going beyond its qualitative role as an indicator of coexistence-like finite-size behavior. In this way, our work goes beyond reaffirming the Potts scenario and resolves the remaining ambiguity through a comprehensive endpoint-scaling analysis.

More broadly, the present work closes a methodological gap in the standard field-mixing framework by extending it beyond its conventional two-relevant-field formulation~\cite{Kwak2015,Mataragkas2023} to endpoint criticality governed by one relevant field and one marginally irrelevant field.
The endpoint can be localized from the mixed-histogram construction and Binder crossings without assuming a specific universality class or its exact critical exponents; candidate scaling forms, critical exponents, and logarithmic corrections enter in the subsequent scaling tests. The construction nevertheless assumes this two-field structure and the availability of the corresponding conjugate operators. It should extend to other endpoints with the same scaling structure, whereas additional scaling fields would require a correspondingly enlarged field-mixing construction. More generally, the underlying principle of disentangling multiple scaling directions may provide a useful framework for more complex multicritical and endpoint problems.
\\

\emph{Acknowledgments}---We thank Rong Yu, Changle Liu, and Yuan Wan for valuable discussions. We also thank Adil A. Gangat for clarifying the interpretation of pseudocriticality in Ref.~\cite{PhysRevB.109.104419}. The authors acknowledge support by the National Natural Science Foundation of China (Grants No. 12404176 and No. 12564022).
\\

\emph{Data availability}---
The data supporting the findings of this study are available in the Zenodo database~\cite{dataset}.

\bibliography{GFMJ1J2}

\clearpage
\newpage
\onecolumngrid
\begin{center}
{\large Supplementary Material for}
$\,$\\ 
$\,$\\ 
\textbf{\large{Relevant--marginal field mixing resolves the four-state Potts endpoint in the square-lattice $J_1$--$J_2$ Ising model}}
$\,$\\
$\,$\\ 
Yihua Sun and Yuchen Fan
\end{center}

\setcounter{figure}{0}                       
\renewcommand{\thefigure}{S\arabic{figure}}  
\renewcommand{\figurename}{Fig.}              

\begin{figure*}[h!]
\includegraphics[angle=0,width=0.86\linewidth]{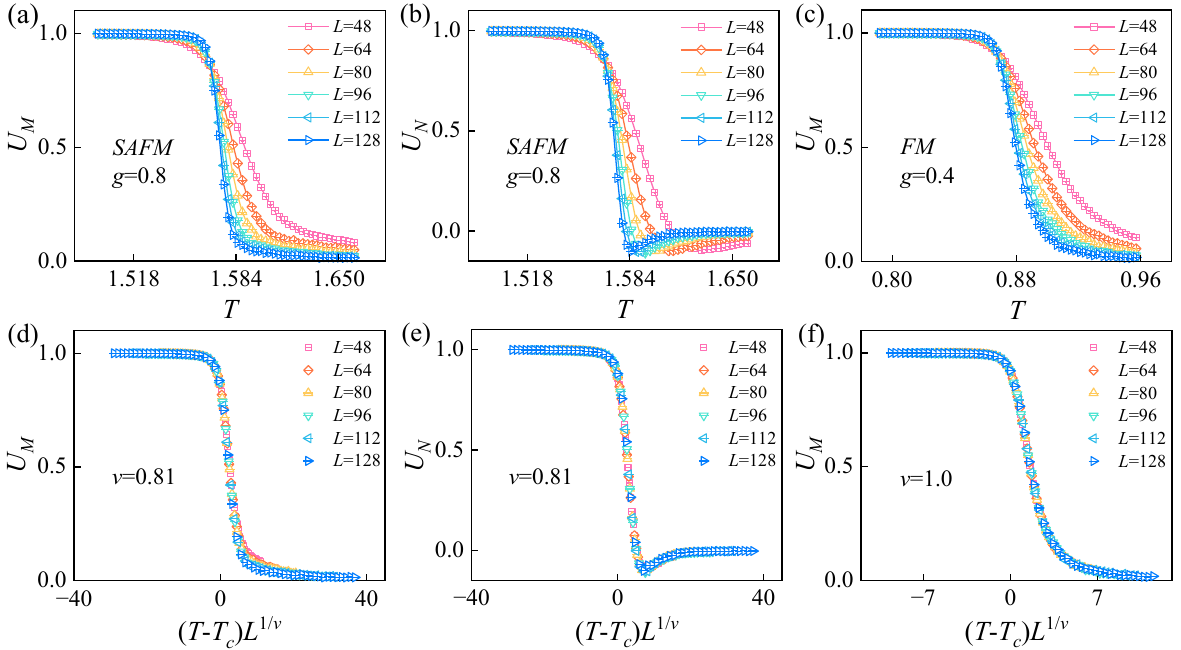}
\caption{
Binder analysis of representative continuous transitions in the square-lattice \(J_1\)--\(J_2\) Ising model.
(a,b) Magnetic and nematic Binder cumulants, \(U_M\) and \(U_N\), for the stripe--paramagnetic transition at \(g=0.8\). The crossings of both Binder ratios locate a common transition temperature, \(T_c=1.5675(4)\).
(c) Binder crossing analysis for the ferromagnetic--paramagnetic transition at \(g=0.4\).
(d,e) Finite-size scaling collapses of \(U_M\) and \(U_N\) in (a,b), giving consistent estimates of the critical exponent \(\nu\).
(f) Finite-size scaling collapse of the data in (c), showing behavior consistent with the two-dimensional Ising universality class.
}
\label{fig:S1}
\end{figure*}

\begin{figure*}[h!]
\includegraphics[angle=0,width=0.38\linewidth]{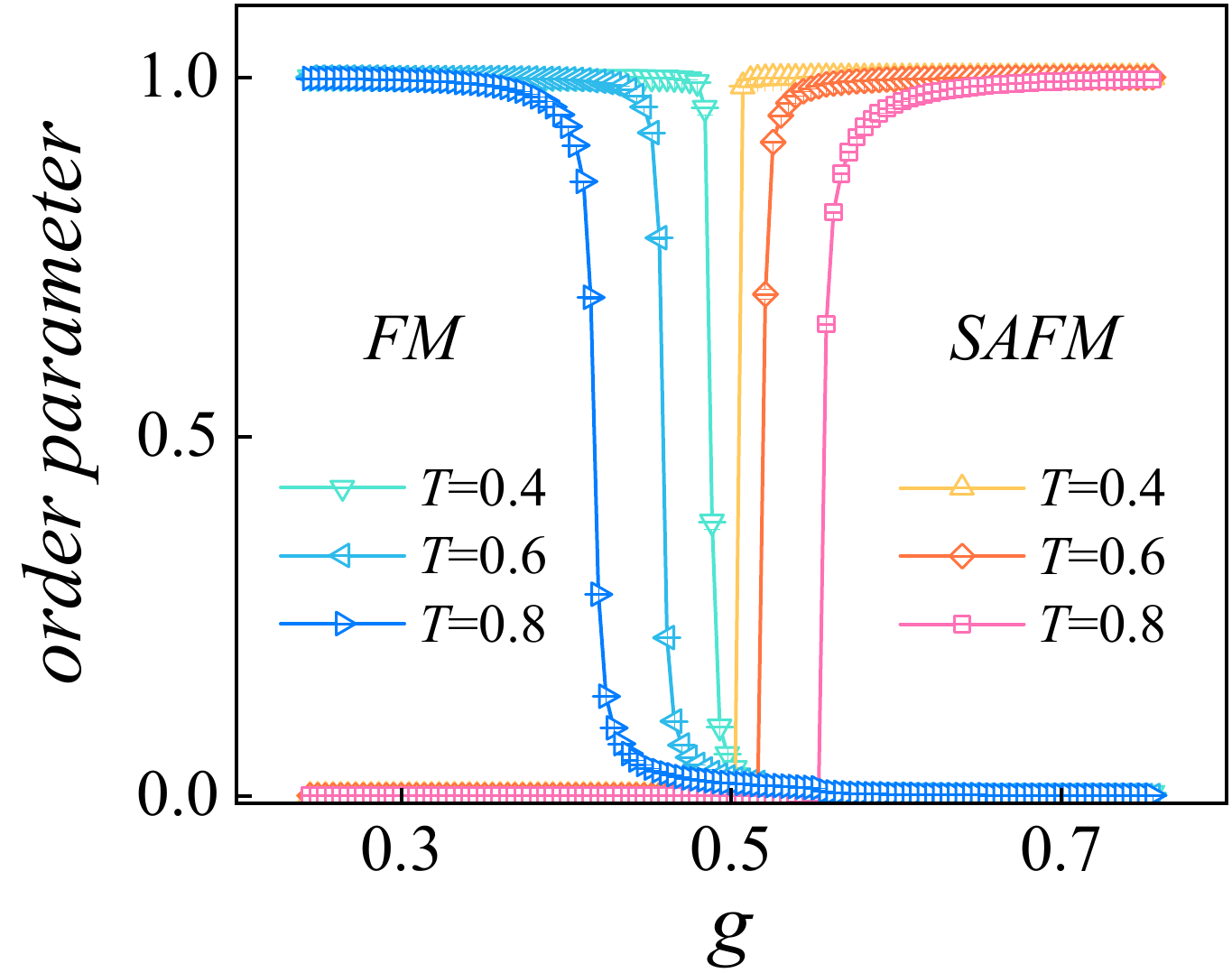}
\caption{
Order-parameter evolution across the phase diagram at fixed temperatures.
We plot the ferromagnetic order parameter
\(M_{\rm FM}=\langle |\frac{1}{L^2}\sum_i \sigma_i| \rangle\)
and the stripe order parameter
\(M_{\rm SAFM}=\langle \sqrt{m_x^2+m_y^2}\rangle\) for system size \(L=96\).
The disappearance of FM order and the onset of stripe order determine the corresponding transition points along each fixed-temperature cut.
As the temperature is lowered, the two transition points approach each other.
Lower temperatures are not included because Monte Carlo equilibration becomes increasingly slow in the first-order regime.
}
\label{fig:S2}
\end{figure*}

\begin{figure*}[h!]
\includegraphics[angle=0,width=0.74\linewidth]{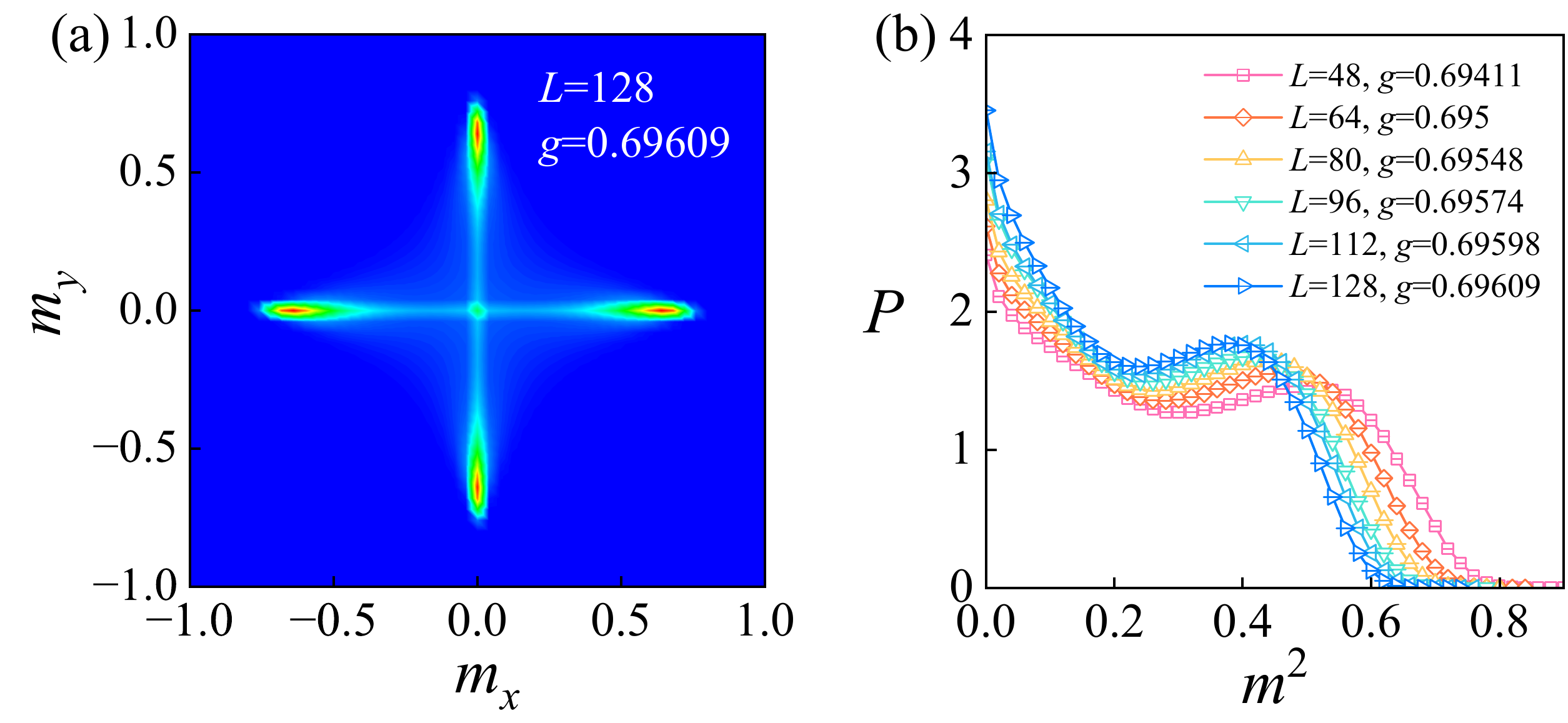}
\caption{
Order-parameter histograms associated with the negative magnetic Binder dip at \(T=T_{\mathrm{ep}}\).
(a) Two-dimensional histograms of the stripe magnetic order parameter \(\mathbf m=(m_x,m_y)\) near the Potts endpoint.
(b) Corresponding probability-density histograms of \(m^2=m_x^2+m_y^2\).
For each system size, the coupling \(g\) is chosen by the equal-weight condition between the ordered and disordered peaks, where the distributions exhibit a pronounced bimodal structure.
The coexistence-like structures visible in both representations provide a direct visualization of the finite-size origin of the negative dip in \(U_M\) shown in Fig. 4(a).
As discussed in the main text, these signatures reflect pseudo-first-order finite-size behavior near the four-state Potts endpoint and should not be interpreted as evidence for a true first-order transition.
}
\label{fig:S3}
\end{figure*}

\end{document}